\documentclass[a4paper,11pt]{article}
\usepackage{jcappub}
\usepackage{lineno}
\usepackage{xcolor}
\usepackage{bm}
\arxivnumber{2606.08368}
\title{\boldmath Reconciling large-scale Lyman-$\alpha$ correlations with the SCRIPT Semi-numerical Model}

\author[1]{Saptarshi Sarkar\note{Corresponding author.}}
\author{and Tirthankar Roy Choudhury}
\affiliation{National Centre for Radio Astrophysics, Tata Institute of Fundamental Research,\\
Pune University Campus, Ganeshkhind, Pune 411007, India}

\emailAdd{sarkar@ncra.tifr.res.in}
\emailAdd{tirth@ncra.tifr.res.in}

\abstract{Recent analyses of high-redshift Lyman-$\alpha$ forest observations have revealed strong correlations on scales exceeding $200\mathrm{\,cMpc}$ at redshift $z = 6$. Reproducing these large-scale correlations has proven challenging for current large-volume reionization simulations. In this work, we investigate these large-scale correlations using mock spectra generated from the extended \texttt{SCRIPT} semi-numerical reionization model. We find that while the fiducial model ensemble systematically predicts smaller correlation lengths than those inferred from the 67 sightlines in the extended XQR-30 sample, 17.5\% of individual mock realizations can naturally reproduce the observed signal. Using a delete-2 jackknife analysis, we demonstrate that the observed large-scale correlation length is disproportionately driven by a rare pair of highly transmissive sightlines associated with high-redshift transmission spikes. By inserting two such highly transmissive sightlines into our mock realizations, the fraction of realizations consistent with the observed redshift evolution and correlation length increases significantly from 17.5\% to 74.1\%. Furthermore, we show that spatial fluctuations in the ionizing mean free path remain an essential physical ingredient for reproducing the observed correlation structure. Our results suggest that the unexpectedly large Lyman-$\alpha$ correlations can be reconciled with existing reionization models when accounting for cosmic variance and the outsized statistical impact of highly transmissive sightlines that occur as prominent outliers within the observational sample.}

\keywords{reionization, Lyman alpha forest}

\begin{document}
\maketitle
\flushbottom

\section{Introduction} \label{sec:introduction}

While the broad chronology of the cosmic reionization is now reasonably well-established~\cite{barkanaBeginningFirstSources2001, choudhuryShortIntroductionReionization2022,chakrabortyReionizationItsSources2026}, the spatial morphology and the specific topology of the ionizing process remain at the forefront of observational studies. The transition from a neutral to an ionized intergalactic medium (IGM) is increasingly viewed as a highly inhomogeneous process, where the distribution of the first luminous sources and the subsequent growth of ionized regions create complex large-scale structures. Characterizing these fluctuations is essential for distinguishing between competing reionization scenarios, particularly those involving the persistence of late-stage neutral islands.

Among the different observational probes of the Epoch of Reionization (EoR), Lyman-$\alpha$ (Ly$\alpha$) spectra from high-redshift quasars have emerged as a particularly powerful tool for studying the state of the IGM during the end stages of the EoR~\cite{gunnDensityNeutralHydrogen1965,beckerReionisationHighRedshiftGalaxies2015}. The large variance observed in the effective optical depth, $\tau_\mathrm{eff}$, of the high-redshift Ly$\alpha$ forest provides compelling evidence for a late and patchy reionization process. Observations reveal that the scatter in $\tau_\mathrm{eff}$ between different quasar sightlines at the same redshift exceeds what would be expected from underlying cosmic density fluctuations alone~\cite{beckerEvidencePatchyHydrogen2015}, a finding robustly confirmed by progressively larger samples of quasar sightlines~\cite{bosmanNewConstraintsLymana2018}. While initial explanations for this excess scatter invoked either large spatial fluctuations in the UV background driven by a short mean free path~\cite{daviesLargeFluctuationsHydrogenionizing2016} or residual temperature fluctuations arising from a patchy and extended reionization process~\cite{daloisioLargeOpacityVariations2015}, recent theoretical work strongly supports scenarios where reionization extends well below redshift $z=6$. In these late reionization models, highly opaque neutral ``islands'' are left behind, which naturally reproduce the longest observed Gunn-Peterson troughs~\cite[e.g.,][]{kulkarniLargeLyOpacity2019,keatingLongTroughsLymana2020,nasirObservingTailReionization2020}. This picture is further supported by computationally efficient semi-numerical models that use fluctuations in the Ly$\alpha$ forest opacity to place quantitative constraints on the end of reionization, indicating that it likely concluded only by $z \lesssim 5.3$--5.6~\cite[e.g.,][]{choudhuryStudyingLymanOptical2021,qinReionizationGalaxyInference2021,qinPercentlevelTimingReionisation2025}. In addition, recent observational datasets, such as the 67 high signal-to-noise quasar sightlines in the extended XQR-30 sample~\cite{dodoricoXQR30UltimateXSHOOTER2023}, have ruled out traditional models assuming a spatially uniform UV background at $z \gtrsim 5.4$, providing strong evidence that hydrogen reionization remained incomplete until as late as $z \approx 5.3$~\cite{bosmanHydrogenReionizationEnds2022}. Complementary observables, including dark-gap statistics and dark-pixel fractions, further support scenarios in which substantial neutral islands and large-scale ionization fluctuations persist down to $z \sim 5.3$--5.5~\cite[e.g.,][]{songailaApproachingReionizationEvolution2002,galleraniConstrainingReionizationHistory2006,fanConstrainingEvolutionIonizing2006,galleraniGlimpsingHighredshiftNeutral2008,zhuChasingTailCosmic2021,maityStudyingDarkGaps2026,mcgreerFirstNearlyModelindependent2011,mcgreerModelindependentEvidenceFavour2015,daviesUpdatedDarkPixel2026}.

More recently, the large-scale correlation properties of the Ly$\alpha$ forest have emerged as a powerful probe of the late stages of reionization. Unlike one-point statistics such as the distribution of effective optical depths or dark-gap lengths, correlation measurements are directly sensitive to the coherence scale of fluctuations in the ionization and thermal state of the IGM, and therefore probe the large-scale morphology of reionization. Using the extended XQR-30 quasar sample~\cite{dodoricoXQR30UltimateXSHOOTER2023}, a recent study by Spina et al.~\cite{spinaMeasuringIntergalacticMedium2026} has reported strong correlations in the transmitted flux extending over scales of $> 200\mathrm{\,cMpc}$ at $z = 6$, with the correlation length increasing rapidly with redshift. However, current large-volume simulations, including models incorporating fluctuating UV backgrounds and residual neutral islands, have been unable to reproduce the observed large-scale correlations~\cite{spinaMeasuringIntergalacticMedium2026, maityStudyingDarkGaps2026}. This discrepancy raises the question of whether the observed correlations reflect limitations in current modeling assumptions or can instead arise from rare statistical fluctuations associated with the finite observed sample, and therefore constitutes an important challenge for existing models of the high-redshift IGM. In this work, we investigate whether such large-scale Ly$\alpha$ forest correlations can be understood and reproduced within the framework of the extended \texttt{SCRIPT} semi-numerical reionization model~\cite{choudhuryPhotonNumberConservation2018,choudhuryCapturingSmallscaleReionization2025}. 

The remainder of this paper is organized as follows: in section~\ref{sec:observations_simulations}, we describe the observational data and the methodology used to generate mock realizations. In section~\ref{sec:correlation_analysis}, we present our correlation analysis framework, including the computation and modeling of the correlation matrix, and the inference of the model parameters from it. The results are presented in section~\ref{sec:results}, and we conclude in section~\ref{sec:summary}. The cosmological parameters used in this work are $\Omega_m = 0.308,\,\Omega_\Lambda=1-\Omega_m,\,\Omega_b=0.0482,\,h=0.678,\,\sigma_8=0.829,\text{ and }n_s=0.961$~\cite{adePlanck2015Results2016}.

\section{Observations and simulations} \label{sec:observations_simulations}

In this section, we describe the observed Ly$\alpha$ spectra used in this work and the methodology adopted to generate mock realizations of the spectra.

\subsection{Observed spectra} \label{subsec:observed_spectra}

In this work we use the mean transmitted flux measurements compiled by Bosman et al. (2022)\footnote{\url{https://academic.oup.com/mnras/article/514/1/55/6598046\#supplementary-data}}~\cite{bosmanHydrogenReionizationEnds2022}, based on sightlines towards 67 quasars at $z > 5.5$ with signal-to-noise ratio $\ge 10$ per $\leq 15\,\mathrm{km\,s^{-1}}$ spectral pixel. These are the same sightlines used in the analysis of Spina et al. (2026)~\cite{spinaMeasuringIntergalacticMedium2026}. The data are provided in redshift bins of $\Delta z = 0.05$ together with asymmetric uncertainties. Following ref.~\cite{spinaMeasuringIntergalacticMedium2026} we retain both detections and upper limits (non-detections) in our sample. By adopting these published measurements directly, we ensure that our study is grounded in a standardized data reduction framework that includes established protocols for signal-to-noise thresholds, spectral pixel binning, and continuum reconstruction. While the core processing is preserved from the original sample, we use these measurements to construct the observed correlation matrix and to match the redshift coverage and noise properties of the mock realizations.

In the published sample~\cite{bosmanHydrogenReionizationEnds2022}, rest-frame wavelengths of $\lambda \le 1026$ \r{A} and $\lambda \ge 1185$ \r{A} are masked to account for Lyman-$\beta$ contamination and quasar proximity effect, respectively. In addition, damped Ly$\alpha$ systems identified along quasar sightlines are also masked. Further details on the sample and the data-cleaning procedures can be found in ref.~\cite{bosmanHydrogenReionizationEnds2022}. For each quasar sightline, we further compute the minimum usable redshift by requiring a separation of $\Delta v = 5000\,\mathrm{km\,s^{-1}}$ from possible O\,VI associated absorption. All redshift bins below this limit are excluded from the analysis to avoid artificial correlations arising from O\,VI contamination.

\subsection{Simulated spectra} \label{subsec:simulations}

\subsubsection{The reionization model} \label{subsubsec:reionization_model}

The reionization model used in this work is based on the semi-numerical, photon-conserving framework \texttt{SCRIPT}~\cite{choudhuryPhotonNumberConservation2018, maityProbingThermalHistory2022}, which predicts the ionization state of the Universe within a cosmologically representative simulation volume. A key feature of the framework is its explicit conservation of ionizing photons, yielding numerically convergent results as the resolution of the ionization maps is varied. We specifically employ the extended fourteen-parameter version of \texttt{SCRIPT}~\cite{choudhuryCapturingSmallscaleReionization2025}, which incorporates several physical processes relevant to the late stages of reionization, including inhomogeneous recombinations, spatially varying mean free paths of ionizing photons, and fluctuations in the photoionization rate. In each grid cell of the simulation volume, inhomogeneous recombinations are modeled using a density-dependent clumping factor derived from a subgrid conditional density distribution. The ionizing mean free path, $\lambda_\mathrm{mfp}$, the clumping factor, and the photoionization background are computed self-consistently through their dependence on the local self-shielding threshold and ionizing emissivity, using the same subgrid density distribution. This approach ensures that these quantities evolve differently across grid cells, thereby self-consistently capturing their spatial fluctuations. The model also follows the thermal evolution of the IGM and incorporates radiative feedback on low-mass galaxies within ionized regions. Together, these ingredients enable the model to capture the large-scale inhomogeneities expected during cosmic reionization.

The extended framework contains fourteen free parameters describing the ionizing source population and the properties of the IGM, including parameters governing the subgrid density distribution and thermal evolution. A more detailed description of the model and its parameters can be found in ref.~\cite{choudhuryCapturingSmallscaleReionization2025}. In this work, we adopt a fiducial parameter set previously shown to satisfy a broad range of observational constraints, including galaxy UV luminosity functions, the CMB optical depth, the hydrogen photoionization rate, the IGM temperature, and the mean free path of ionizing photons~\cite{choudhuryCapturingSmallscaleReionization2025}. Our goal here is not to fit for all the model parameters, but to test whether a fiducial model already consistent with existing constraints can reproduce the observed correlation statistics. For the purpose of this work, following ref.~\cite{spinaMeasuringIntergalacticMedium2026}, we use a simulation box size of $1\,h^{-1}\mathrm{\,cGpc}$ with $256^3$ grid cells. This volume is sufficient to capture the large scales relevant for the observed correlations.

\subsubsection{Lyman-\texorpdfstring{$\alpha$}{alpha} optical depth} \label{subsubsec:optical_depth}

As our aim is to generate mock realizations of the Ly$\alpha$ spectra, we construct lightcones of the Ly$\alpha$ optical depth, $\tau_\alpha$. For this purpose, we use simulation boxes spanning redshifts $z=6.1$ to $5$ at intervals of $\Delta z=0.1$, and interpolate to obtain the intermediate redshift slices. For each box, $\tau_\alpha$ is computed in each grid cell using the fluctuating Gunn-Peterson approximation (FGPA) based formalism outlined in ref.~\cite{choudhuryStudyingLymanOptical2021}. Under this approximation, the Ly$\alpha$ optical depth in the $i$th cell is given by~\cite{choudhuryStudyingLymanOptical2021, choudhuryCapturingSmallscaleReionization2025}
\begin{align} \label{eq:gunn_peterson}
    \tau_{\alpha,i}=\kappa_{\mathrm{res}}\frac{\pi e^2}{m_e c}f_\alpha\lambda_\alpha\frac{(1+z)^6}{H(z)}\chi_{\mathrm{He}}\frac{\alpha_B(T_{\mathrm{HII},i})n^2_{\mathrm{H},i}}{\Gamma_{\mathrm{HI},i}},
\end{align}
where $f_\alpha$ and $\lambda_\alpha$ are the oscillator strength and rest-frame wavelength of the Ly$\alpha$ transition. The factor $\chi_{\mathrm{He}}$ accounts for the contribution of singly ionized helium to the free-electron density. Further, $\alpha_B(T_{\mathrm{HII},i})$ is the case-B recombination rate evaluated at the ionized-gas temperature $T_{\mathrm{HII},i}$, $n_{H,i}$ is the hydrogen number density, and $\Gamma_{\mathrm{HI},i}$ is the hydrogen photoionization rate in the $i$th cell. The parameter $\kappa_{\mathrm{res}}$ is an overall normalization factor that accounts for small-scale density and velocity fluctuations that remain unresolved in our simulations. The $\kappa_{\mathrm{res}}$ values are redshift-dependent and are chosen to reproduce the observed distribution of effective Ly$\alpha$ forest optical depths~\cite{choudhuryStudyingLymanOptical2021, choudhuryCapturingSmallscaleReionization2025}; the corresponding validation plots are presented in appendix~\ref{apdx:validation_taueffs}. 

The above expression assumes photoionization equilibrium in ionized regions and is therefore applicable only for fully ionized cells. For cells containing neutral regions, the optical depth in the ionized portion of the cell is given by $x_{\mathrm{HII},i}\tau_{\alpha,i}$~\cite{choudhuryCapturingSmallscaleReionization2025}, where $x_{\mathrm{HII},i}$ is the hydrogen ionization fraction, and $\tau_{\alpha,i}$ is determined by eq.~\ref{eq:gunn_peterson}. The optical depth in the neutral portion is assumed to be effectively infinite, resulting in zero transmitted flux.

It is important to note that, while our approach is based on the FGPA-based formalism of ref.~\cite{choudhuryStudyingLymanOptical2021}, Spina et al. (2026)~\cite{spinaMeasuringIntergalacticMedium2026} adopt a modified FGPA relation calibrated using high-resolution hydrodynamical Nyx simulations~\cite{maityEfficientModelingLymana2026}. The differing treatments can therefore lead to quantitative differences in the resulting Ly$\alpha$ opacity fluctuations.

\subsubsection{Mock realizations} \label{subsubsec:mock_realizations}

To generate each mock realization, we randomly extract 67 skewers from the Ly$\alpha$ optical depth lightcone along the line of sight, matching the size of the observed data. We generate 1000 such mock realizations. We note that our lightcones contain only $256^2 (= 65536)$ distinct skewers. Consequently, the same skewer may appear in different realizations. However, we ensure that no skewer is repeated within any individual realization.

\begin{figure}[htbp]
\centering
\includegraphics[width=\textwidth]{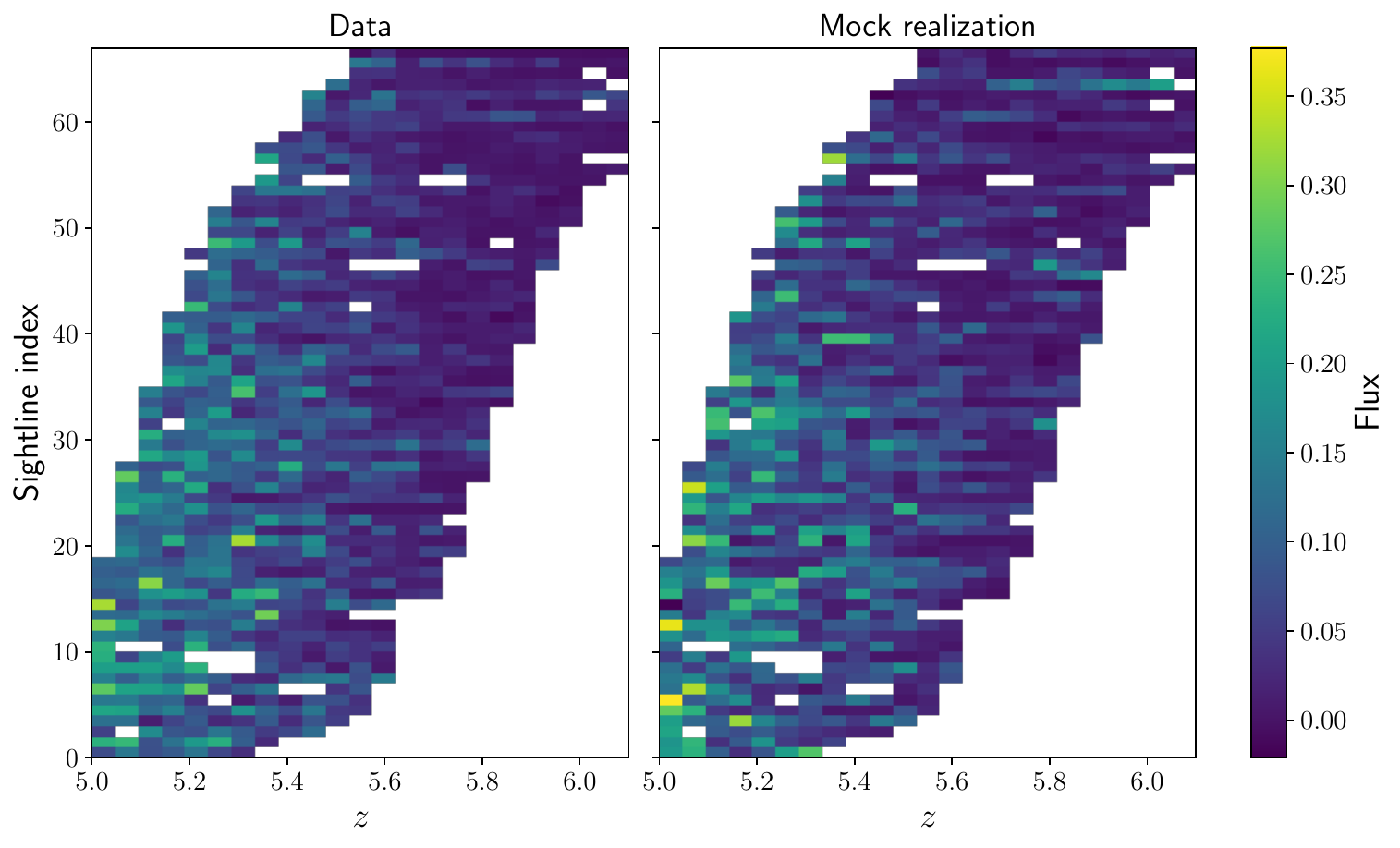}
\caption{Comparison of the observed Ly$\alpha$ spectra (\textit{left panel}) with a mock realization from the fiducial model (\textit{right panel}). The mock realizations are constructed using the same redshift binning, masking pattern, and observational noise properties as the data. White regions indicate masked or unavailable bins.\label{fig:spectra_comparison}}
\end{figure}

Since the observable of interest is the transmitted flux rather than the optical depth, for each skewer, we compute the transmitted flux in a given cell $i$ as
\begin{align} \label{eq:flux}
    F_i = x_{\mathrm{HII},i}\exp\!\left(-x_{\mathrm{HII},i}\tau_{\alpha,i}\right).
\end{align}
To make each mock realization as representative of the observations as possible, we adopt the same redshift bins as the observed data by binning the skewers into redshift intervals of $\Delta z = 0.05$ and computing the mean transmitted flux in each bin. We additionally apply the same masking pattern as in the observed spectra, removing all bins that are excluded from the observed data. We also add observational noise in each redshift bin using the corresponding measured uncertainties. To account for the asymmetric observational error bars, the noise realizations are drawn from a piecewise Gaussian distribution with separate upper and lower standard deviations. A visual comparison between the observed spectra and a mock realization from the fiducial model is shown in figure~\ref{fig:spectra_comparison}, illustrating that the mock reproduces the redshift coverage, masking pattern, and overall flux distribution of the data.

\section{Correlation analysis} \label{sec:correlation_analysis}

In this section, we describe the statistical framework used to model and infer the Ly$\alpha$ forest correlation parameters from both the observed data and the mock realizations.

\subsection{Correlation matrix} \label{subsec:correlation_matrix}

Following ref.~\cite{spinaMeasuringIntergalacticMedium2026}, we measure correlations in the Ly$\alpha$ forest spectra using a Pearson correlation based formalism. Let $S_i$ denote the transmitted-flux spectrum of the $i$th sightline. The mean spectrum is then defined as
\begin{align}
    \bar{S} = \frac{1}{N}\sum_{i=1}^{N} S_i,
\end{align}
where $N=67$ is the total number of sightlines in the sample. We next compute the covariance between redshift bins $j$ and $k$ as
\begin{align}
    C_{jk} = \frac{1}{N_{jk}-1}\sum_{i\in\mathcal{V}_{jk}}
\left(S_i^{(j)}-\bar{S}^{(j)}\right)
\left(S_i^{(k)}-\bar{S}^{(k)}\right),
\end{align}
where $\mathcal{V}_{jk}$ is the set of $N_{jk}$ sightlines with valid measurements in both bins $j$ and $k$, $S_i^{(j)}$ is the transmitted flux of the $i$th sightline in bin $j$, and $\bar{S}^{(j)}$ is the value of the mean spectrum in that bin. Finally, the correlation matrix is obtained as
\begin{align}
    \mathcal{D}_{jk} = \frac{C_{jk}}{\sqrt{C_{jj}C_{kk}}}.
\end{align}

We note that $N_{jk}$ is not necessarily equal to 67 for all pairs of redshift bins. The redshift interval over which the Ly$\alpha$ forest is usable depends on the redshift and masking pattern of each quasar sightline, and therefore not all quasars contribute to every bin. In particular, bins near the boundaries of the redshift range are sampled by fewer sightlines than those in the central bins (see figure~\ref{fig:spectra_comparison}). If the same set of quasars contributed to both bins $j$ and $k$, then $\mathcal{D}_{jk}$ would reduce to the standard Pearson correlation coefficient and would therefore be bounded within $[-1,1]$. As a result, the quantity $\mathcal{D}_{jk}$ defined above is not guaranteed to lie strictly within the interval $[-1,1]$. Nevertheless, it serves as a useful estimator of the underlying correlation structure of the data.

We estimate the standard deviation $\sigma_{jk}$ of the correlation matrix $\mathcal{D}_{jk}$ separately for the observed and mock realizations. For the observed spectra, the flux values are randomly permuted across sightlines within each redshift bin, and this procedure is repeated 1000 times. A correlation matrix is computed for each shuffled realization, and the standard deviation over the resulting matrices is taken as the uncertainty estimate. For the mock realizations, we estimate the uncertainty by taking the standard deviation of the correlation matrices across the 1000 realizations. The resulting uncertainty is then adopted for all individual mock realizations.

We also note that ref.~\cite{spinaMeasuringIntergalacticMedium2026} additionally modeled the effect of continuum-reconstruction uncertainties using dedicated mock observations. We do not include this extra step here, but instead use the published mean transmitted-flux measurements directly. The resulting inferred correlation parameters remain in close agreement with their published constraints, with only a slight shift in the correlation amplitude (see appendix~\ref{apdx:validation_analysis_pipeline}), indicating that this simplification does not affect our main conclusions.

\subsection{Modeling and inference} \label{subsec:modeling_inference}

We model and infer the correlation parameters using the parametrization introduced in ref.~\cite{spinaMeasuringIntergalacticMedium2026}. In that work, this model is referred to as $\mathcal{M}_3$; since we consider only this parametrization here, we refer to it simply as $\mathcal{M}$. The model is given by
\begin{align}
    \mathcal{M}(z_1,z_2) &= A(z_1,z_2)\exp\left[-\frac{1}{2}\left(\frac{\chi(z_1)-\chi(z_2)}{L(z_1,z_2)}\right)^2\right],\\
    A(z_1,z_2) &= A_0 + A_1(z_2-5),\\
    L(z_1,z_2) &= L_0 + L_1(z_2-5),
\end{align}
where $\chi(z)$ is the comoving distance to redshift $z$, $A$ is the correlation amplitude, and $L$ is the correlation length. The free parameters $(A_0, A_1, \log_{10}L_0, \log_{10}L_1)$ are constrained by fitting this model to the measured correlation matrix.

We constrain the free parameters of the model using a Markov chain Monte Carlo (MCMC) analysis, which samples the posterior distribution of the parameter set $\theta$ given the measured data vector $\mathcal{D}$. The posterior distribution, $\mathcal{P}(\theta \mid \mathcal{D})$, is obtained from Bayes' theorem
\begin{equation}
    \mathcal{P}(\theta \mid \mathcal{D}) =
    \frac{\mathcal{L}(\mathcal{D} \mid \theta)\,\pi(\theta)}
    {\mathcal{P}(\mathcal{D})},
\end{equation}
where $\mathcal{L}(\mathcal{D} \mid \theta)$ is the likelihood of the data for a given set of parameters, $\pi(\theta)$ is the prior distribution, and $\mathcal{P}(\mathcal{D})$ is the Bayesian evidence, which serves as a normalization constant. 

We adopt the likelihood
\begin{align}
\mathcal{L}(\mathcal{D}\mid\theta)
= \exp\left[
-\frac{1}{2}
\sum_{j}\sum_{k<j}
\frac{\left(\mathcal{D}_{jk}-\mathcal{M}_{jk}(\theta)\right)^2}{\sigma_{jk}^2}
\right],
\end{align}
where $\mathcal{M}_{jk}(\theta)$ denotes the model evaluated for the pair of redshift bins $(j,k)$, corresponding to $\mathcal{M}(z_1,z_2)$ with $z_1$ and $z_2$ taken as the redshifts of the $j$th and $k$th bins, respectively. The quantity $\sigma_{jk}$ is the estimated standard deviation of $\mathcal{D}_{jk}$. The priors used for all model parameters are specified in table~\ref{tab:priors}.

\begin{table}[htbp]
\centering
\begin{tabular}{ll}
\hline
Parameter & Prior Range \\
\hline
$A_0$ & [$-$10, 10] \\
$A_1$ & [$-$10, 10] \\
$\log_{10} L_0$ & [0, 3] \\
$\log_{10} L_1$ & [0, 3]\\
\hline
\end{tabular}
\caption{Prior ranges for the free model parameters.\label{tab:priors}}
\end{table}

To sample the model parameter space and obtain the posterior distributions, we use the publicly available package \texttt{Cobaya}~\cite{torradoCobayaBayesianAnalysis2019,torradoCobayaCodeBayesian2021}. In particular, we employ its Metropolis-Hastings MCMC sampler~\cite{metropolisEquationStateCalculations1953, lewisEfficientSamplingFast2013}. The analysis is performed with 8 independent parallel chains~\cite{lewisCosmologicalParametersCMB2002, lewisEfficientSamplingFast2013}. Convergence is assessed using the Gelman-Rubin $R-1$ criterion~\cite{gelmanInferenceIterativeSimulation1992}, requiring $R-1 < 0.01$ in two successive checks. After convergence, the initial 30\% of samples from each chain are removed as burn-in, and the remaining samples are used for parameter estimation. The resulting posterior distributions are processed using the \texttt{GetDist} package~\cite{lewisGetDistPythonPackage2019}. Using this framework, we estimate the posterior distributions for the observed sample, as well as for each of the mock realizations. We validate our correlation matrix analysis pipeline in appendix~\ref{apdx:validation_analysis_pipeline} by comparing our results with the published results obtained from the observed data.

\section{Results} \label{sec:results}

\subsection{Fiducial model} \label{subsec:fiducial_model}

\begin{figure}[htbp]
\centering
\includegraphics[width=\textwidth]{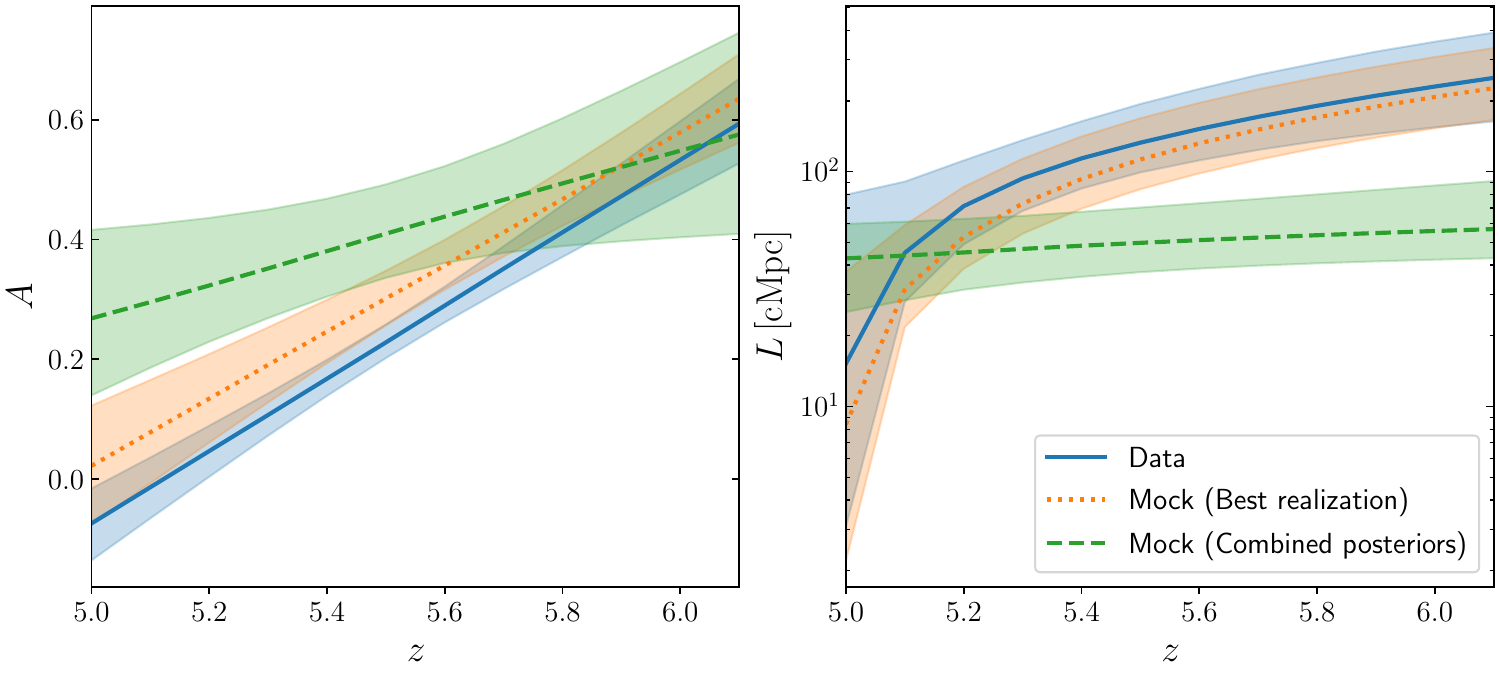}
\caption{Redshift evolution of the correlation amplitude and length. The results for the observed sample are shown in blue in both panels. The green curves show the corresponding results for the mock realizations, obtained by combining the posterior samples from all 1000 realizations, and the orange curves show the result for one representative realization that best matches the data. The central curves represent the 50th percentile of the posterior distribution, while the shaded bands enclose the 16th--84th percentile range. \label{fig:correlations_fiducial}}
\end{figure}

It is important to note that, owing to the finite number of sightlines in the observed sample, cosmic variance must be taken into account when comparing it with simulation results. We therefore compare the correlation parameters inferred from the observations with our full ensemble of mock realizations, rather than with any single realization. Figure~\ref{fig:correlations_fiducial} shows the redshift evolution of the inferred correlation amplitude and length for the observed spectra together with the corresponding results from the mock realizations, obtained by performing an independent MCMC analysis for each of the 1000 realizations and combining the posterior samples. The inferred redshift evolution from the simulations is consistent with that reported in ref.~\cite{spinaMeasuringIntergalacticMedium2026}. In particular, the simulations systematically predict smaller correlation lengths than those inferred from the data.

However, while the ensemble describes the typical behavior of the fiducial model, it does not by itself indicate whether rare realizations can reproduce the large correlation lengths seen in the observations. If such cases occur with non-negligible probability, the observed signal could instead be explained as a consequence of cosmic variance within the fiducial model, rather than as evidence for additional physical effects not captured in the fiducial model. Indeed, when the mock realizations are examined individually, we find a small number of cases whose inferred correlation parameters are consistent with the observations; figure~\ref{fig:correlations_fiducial} shows one realization that most closely matches the data.

To quantify consistency with the data, we use the Bhattacharyya distance~\cite{bhattacharyyaMeasureDivergenceTwo1946}, $D_B$, as a measure of the separation between the posterior distribution of the observed data and those of the mock realizations. Approximating the posteriors to be Gaussian, we use~\cite{kashyapPerfectMarriageMuch2019}
\begin{align} \label{eq:bhattacharyya}
D_B = \frac18(\bm{\mu}_{\mathrm{data}}-\bm{\mu}_{\mathrm{mock}})^T\bm{\Sigma}^{-1}(\bm{\mu}_{\mathrm{data}}-\bm{\mu}_{\mathrm{mock}})
+\frac12\ln\left(\frac{\det\bm{\Sigma}}{\sqrt{\det\bm{\Sigma}_{\mathrm{data}}\det\bm{\Sigma}_{\mathrm{mock}}}}\right),
\end{align}
where $\bm{\mu}_{\mathrm{data}}$ and $\bm{\Sigma}_{\mathrm{data}}$ are the mean vector and covariance matrix computed from the posterior samples of the observed data, while $\bm{\mu}_{\mathrm{mock}}$ and $\bm{\Sigma}_{\mathrm{mock}}$ are obtained in the same way from the posterior samples of a given mock realization. We define $\bm{\Sigma}=\frac12(\bm{\Sigma}_{\mathrm{data}}+\bm{\Sigma}_{\mathrm{mock}})$. The Bhattacharyya distance quantifies the separation between two distributions by accounting for differences in both their mean values and covariance structures. Consequently, mock realizations with small $D_B$ correspond to cases in which both the inferred correlation parameters and their associated uncertainties are similar to those obtained from the observations. We note, however, that equation~\ref{eq:bhattacharyya} assumes the posterior distributions to be Gaussian and therefore does not capture possible non-Gaussian features in the posteriors. The distribution of $D_B$ for the mock realizations of the fiducial model is shown in the left panel of figure~\ref{fig:bhattacharyya} with solid lines. The figure shows a broad distribution of $D_B$ values across the mock realizations, with a tail extending toward low values of $D_B$. This suggests that while the fiducial model does not typically reproduce the observed correlation signal, a small number of realizations have posterior distributions broadly consistent with the observations.

To understand why a subset of realizations yields correlation parameters consistent with the observations, we examined the high-redshift transmitted-flux properties of the mock realizations. Although largely stochastic, we find that the Bhattacharyya distance, $D_B$, is weakly anti-correlated with the mean flux of the most transmissive sightlines and weakly correlated with the mean flux of the least transmissive sightlines, where the ranking is based on their mean flux at $z>5.5$. Consistent with this, $D_B$ is also weakly anti-correlated with the sightline-to-sightline scatter in the mean transmitted flux, suggesting that realizations with stronger flux fluctuations are more likely to reproduce the observed Ly$\alpha$ correlation signal. This motivates a closer examination of the role played by rare sightlines in the observed sample.

\subsection{Sensitivity to individual sightlines} \label{subsec:outlier_analysis}

To further investigate why only a fraction of mock realizations yield correlation parameters consistent with the observations, we examine whether the inferred parameters are highly influenced by a small subset of sightlines in the observed sample. If the measured signal is highly sensitive to the inclusion of particular quasar sightlines, then removing those could lead to a significant change in the recovered correlation parameters.

To test this, we perform a delete-$d$ jackknife analysis for $d=1$ and 2. For each case, we repeat the full MCMC analysis after excluding $d$ sightlines from the observed sample, considering all possible combinations of omitted spectra. Thus, for $d=1$ and $2$, the number of independent MCMC runs is ${}^{67}C_{1}=67$ and ${}^{67}C_{2}=2211$, respectively. We restrict our analysis to a maximum of $d=2$ as exploring higher-order combinations (e.g., ${}^{67}C_{3} = 47905$ independent MCMC runs for $d=3$) becomes computationally prohibitive. Furthermore, as shown below, the $d=2$ case proves sufficient to isolate the dominant outlier sightlines. 

\begin{figure}[htbp]
\centering
\includegraphics[width=\textwidth]{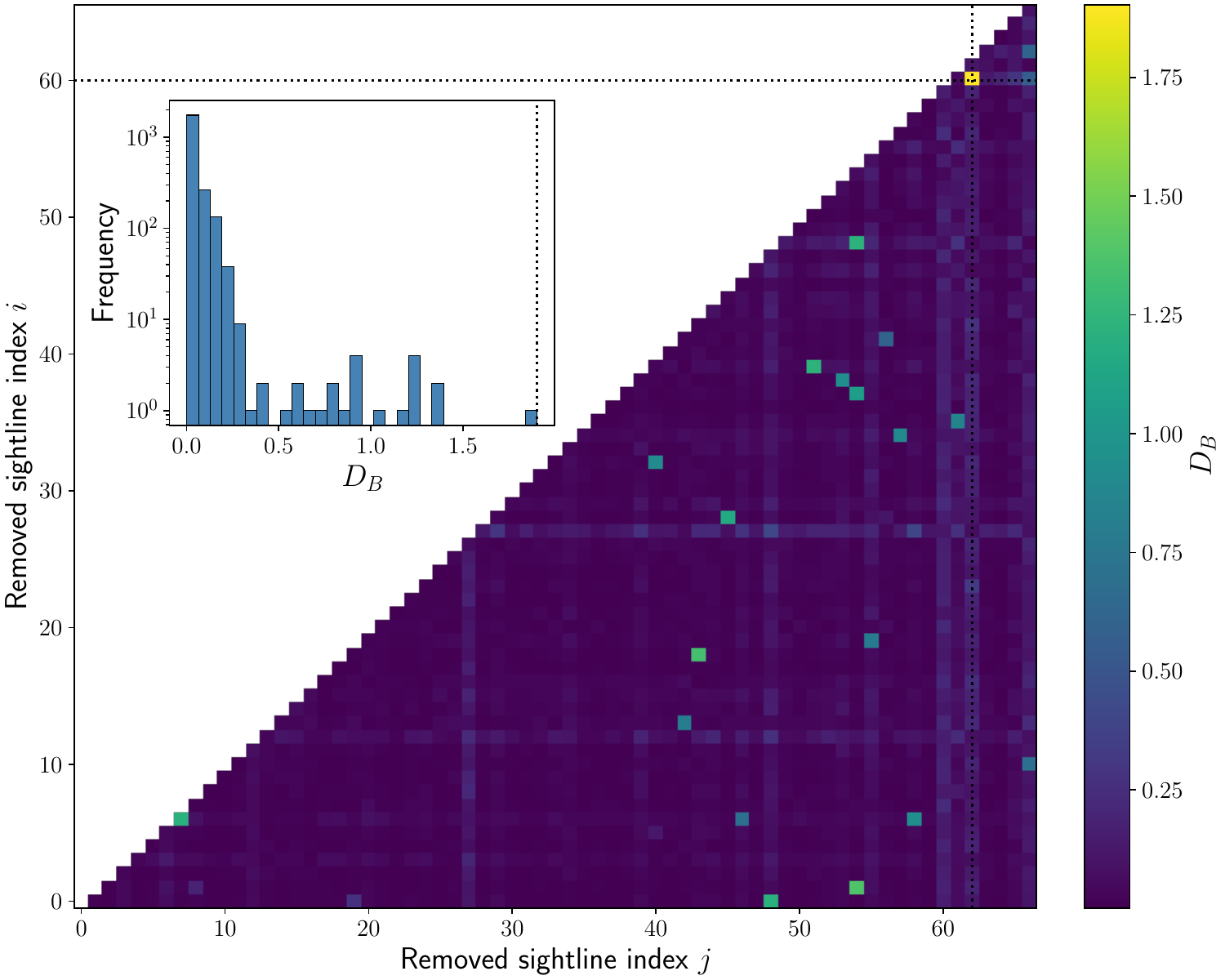}
\caption{Bhattacharyya distance ($D_B$) between the posterior inferred from the full observational dataset and that obtained after removing the pair of sightlines $(i,j)$ in the delete-2 jackknife analysis. The heatmap shows that removing the pair $(60,62)$, marked by the dotted lines, yields the largest distance, $D_B=1.90$, demonstrating that these two sightlines have a disproportionate influence on the inferred correlation parameters. The inset shows the distribution of $D_B$ values over all tested pairs. \label{fig:jackknife_corner}}
\end{figure}

\begin{figure}[htbp]
\centering
\includegraphics[width=0.75\textwidth]{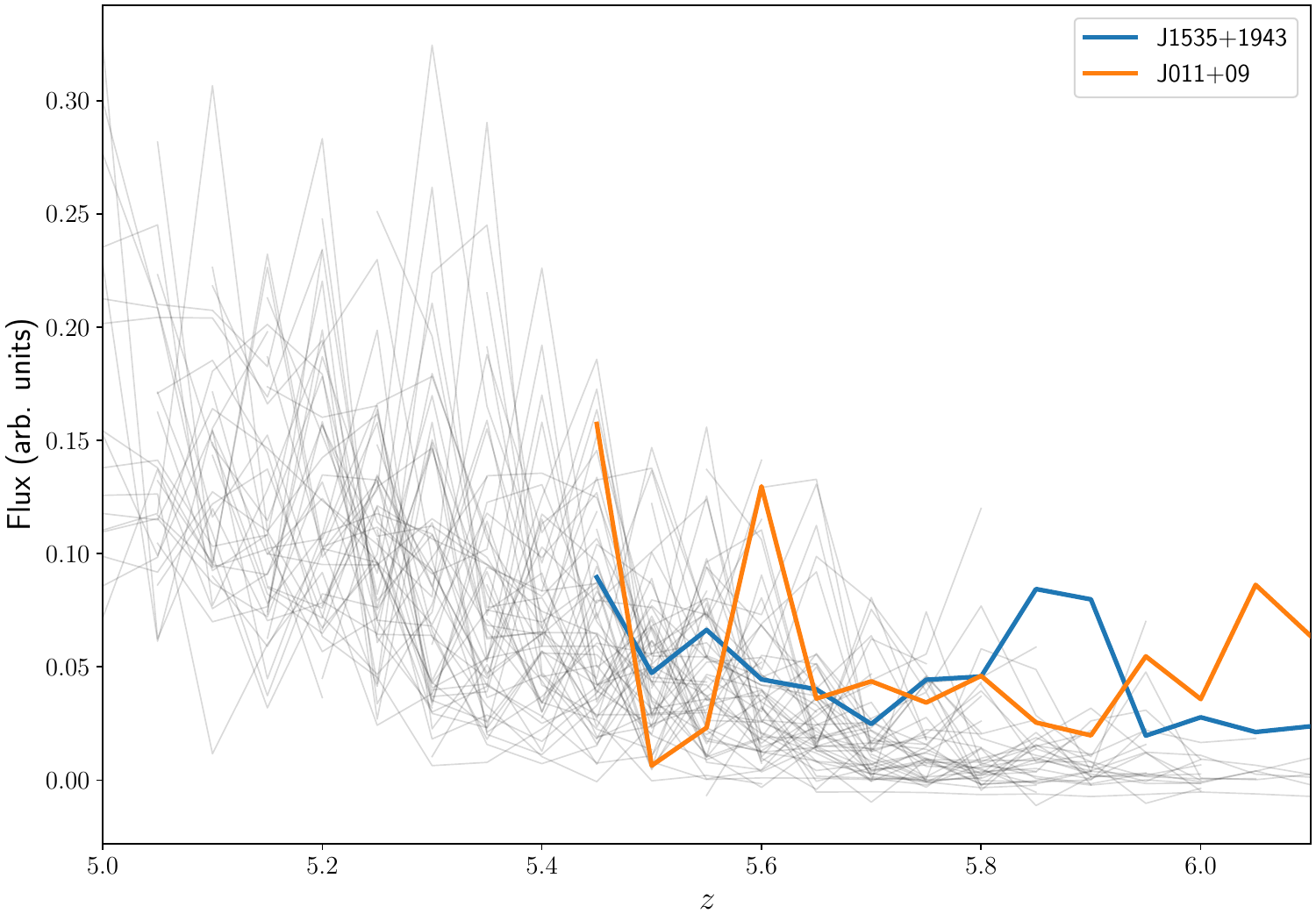}
\caption{Observed spectra of the 67 quasar sightlines in the sample. The two sightlines identified by the delete-2 jackknife analysis, J1535$+$1943 and PSO J011$+$09, are highlighted. \label{fig:observed_spectra}}
\end{figure}

\begin{figure}[htbp]
\centering
\includegraphics[width=\textwidth]{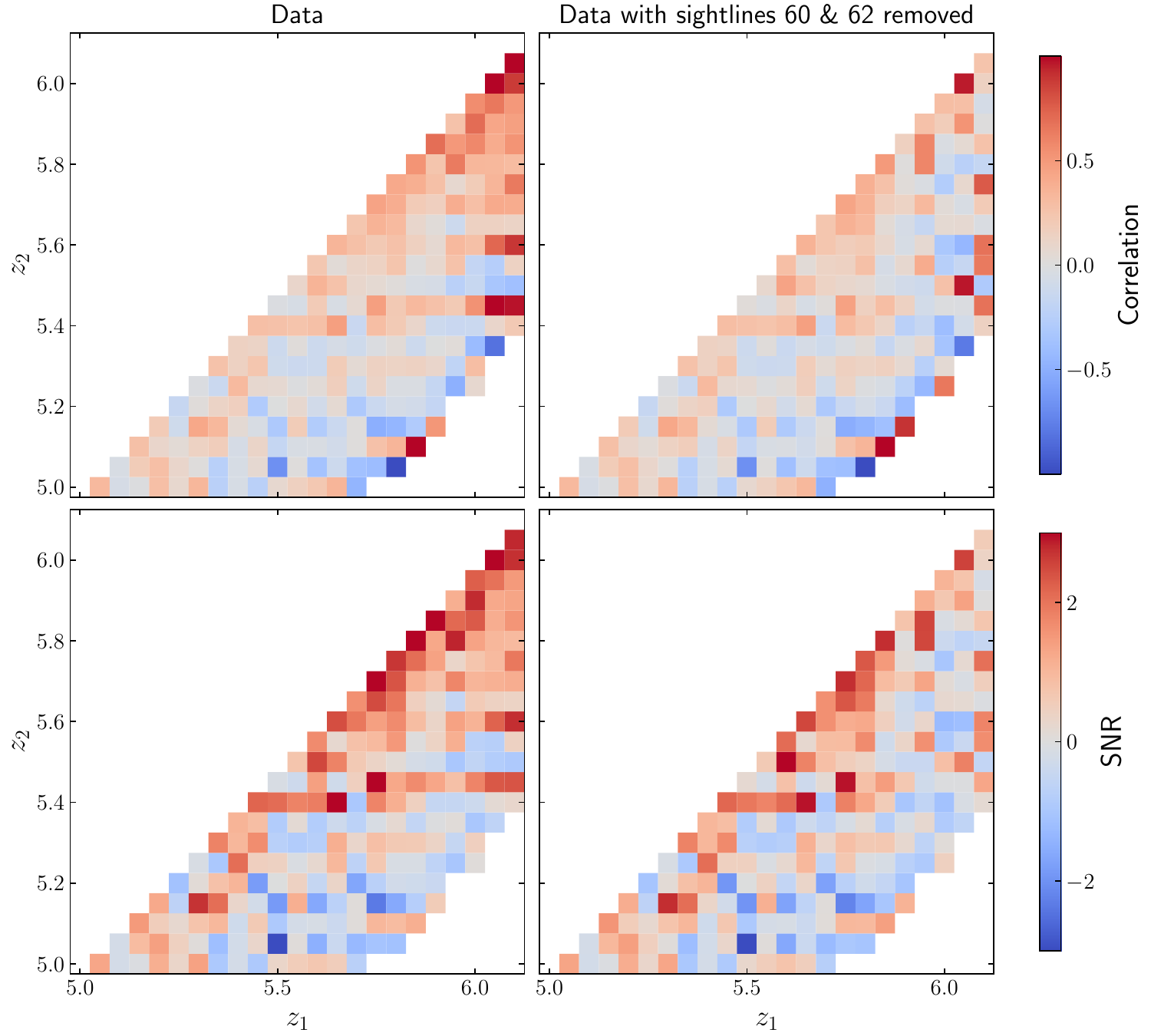}
\caption{Comparison of the correlation matrix $\mathcal{D}_{jk}$, and the SNR matrix $\mathcal{D}_{jk}/\sigma_{jk}$ computed using the full observed sample from \cite{bosmanHydrogenReionizationEnds2022} (\textit{left panel}) with the ones computed after removing sightlines 60 \& 62 (\textit{right panel}). \label{fig:corr_matrices_comp_cropped_2}}
\end{figure}

\begin{figure}[htbp]
\centering
\includegraphics[width=\textwidth]{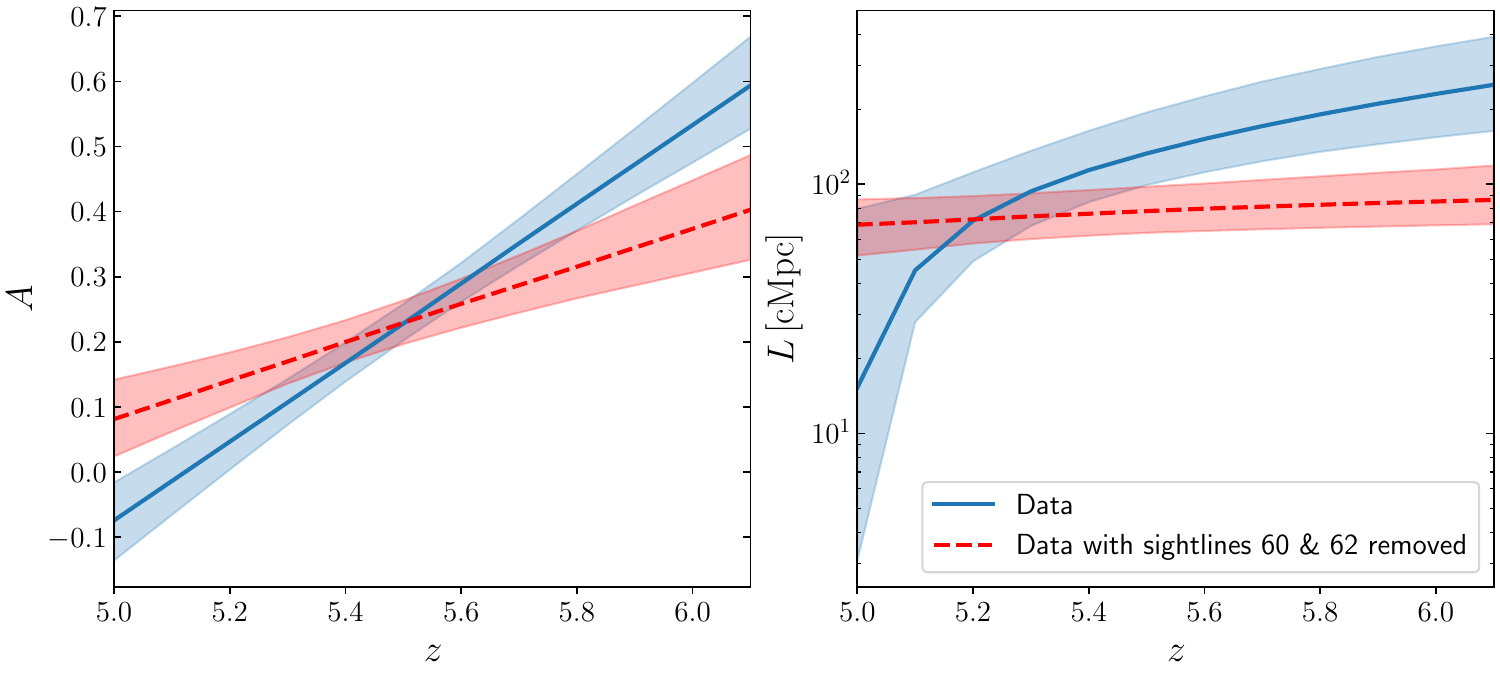}
\caption{Redshift evolution of the correlation amplitude and length. The results for the full observed sample are shown in blue, while those obtained after removing the sightlines 60 \& 62 are shown in red. Formatting is the same as in figure~\ref{fig:correlations_fiducial}. \label{fig:correlations_jackknife2}}
\end{figure}

In our analysis, we find no significant change in the inferred correlation parameters under the delete-1 jackknife analysis, indicating that the observed signal is not driven by any single sightline. However, the delete-2 jackknife analysis reveals a pair of sightlines whose removal leads to a substantial decrease in the inferred correlation length. Figure~\ref{fig:jackknife_corner} shows the $D_B$ values between the posterior from the full sample and that obtained after removing each pair of sightlines. In particular, the pair (60, 62) yields the largest value, $D_B=1.90$, while all other pairs are found to be consistent with the posterior from the full sample. These two sightlines correspond to the quasars J1535$+$1943 and PSO J011$+$09. Interestingly, both of these are highly transmissive at high redshifts, and were also identified in ref.~\cite{bosmanHydrogenReionizationEnds2022} as the most transmissive sightlines in the sample at $z=5.9$ and $z=6.1$, respectively. Figure~\ref{fig:observed_spectra} shows the observed spectra of all 67 sightlines, with these two sightlines highlighted. This high transmissivity has been associated with strong high-redshift transmission spikes along these sightlines. Numerical simulations suggest that such spikes originate from underdense, highly ionized regions of the intergalactic medium~\cite[e.g.,][]{gnedinCosmicReionizationComputers2017,chardinTaleSevenNarrow2018,kakiichiRoleGalaxiesAGN2018,garaldiConstrainingTailEnd2019,nasirObservingTailReionization2020,gaikwadProbingThermalState2020}. Figure~\ref{fig:corr_matrices_comp_cropped_2} compares the correlation $\mathcal{D}_{jk}$ and signal-to-noise ratio (SNR) $\mathcal{D}_{jk}/\sigma_{jk}$ matrices obtained from the full observed sample with those after removing these two sightlines. The corresponding evolution of the inferred correlation amplitude and correlation length is shown in Figure~\ref{fig:correlations_jackknife2}. This demonstrates that the inferred large-scale correlation signal can be strongly influenced by a rare pair of highly transmissive sightlines whose inclusion contributes substantially to the observed excess correlation length.

We also investigated the other sightline pairs with large $D_B$ values in the delete-2 jackknife analysis. In all cases, the inferred correlation amplitude and correlation length remain consistent with those obtained from the full dataset, indicating that these pairs do not significantly influence our results. Although these pairs exhibit relatively large $D_B$ values, this is primarily due to changes in the posterior covariance. In contrast, the pair $(60,62)$ produces a substantial shift in the posterior mean, accounting for its unique impact on the inferred correlation length.

\subsection{Inclusion of highly transmissive sightlines} \label{subsec:biased_results}

Motivated by the jackknife results, we next test whether the inclusion of highly transmissive sightlines can increase the likelihood of reproducing the observed correlations in the mock realizations. To do so, we modify the original set of mock realizations by inserting two highly transmissive sightlines at the same sightline indices (i.e., 60 \& 62) as in the observed sample. This ensures that the inserted sightlines inherit the same redshift bins, masking pattern, and noise properties as the corresponding observed sightlines. We then examine whether this procedure increases the fraction of mock realizations whose inferred correlation parameters are consistent with the observations.

\begin{figure}[htbp]
\centering
\includegraphics[width=\textwidth]{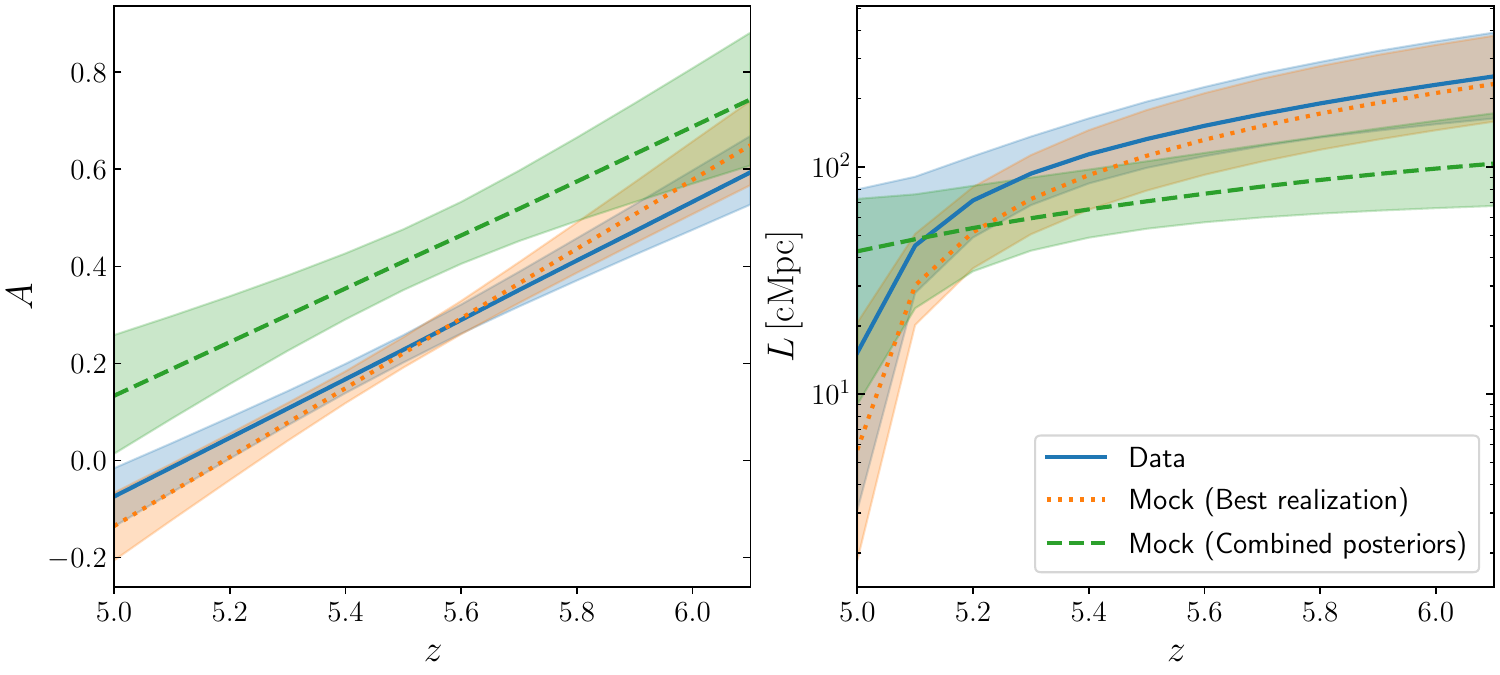}
\caption{Redshift evolution of the correlation amplitude and length. The mock realizations have been modified to include two sightlines that are highly transmissive at $z>5.5$. Formatting is the same as in figure~\ref{fig:correlations_fiducial}. \label{fig:correlations_biased_two}}
\end{figure}

For each of the mock realizations, we randomly select two skewers from the subset with the highest mean transmitted flux at $z>5.5$ (top 1\% of all $256^2$ available skewers). These selected skewers are then processed using the same redshift binning and observational noise properties as the sightlines they replace, and inserted into the realization at the corresponding indices. We additionally ensure that no inserted skewer is already present in that realization. We then repeat the previous analysis by running an independent MCMC analysis for each of the 1000 modified mock realizations and combining the resulting posterior samples. Figure~\ref{fig:correlations_biased_two} shows the inferred correlation parameters as a function of redshift. The close agreement between the combined distribution and the observations, particularly for the correlation length, indicates that a large fraction of the modified mock realizations now reproduce the observed signal; the realization that best matches the data (smallest $D_B$) is shown in figure~\ref{fig:correlations_biased_two} in orange. 

\begin{figure}[htbp]
\centering
\includegraphics[width=\textwidth]{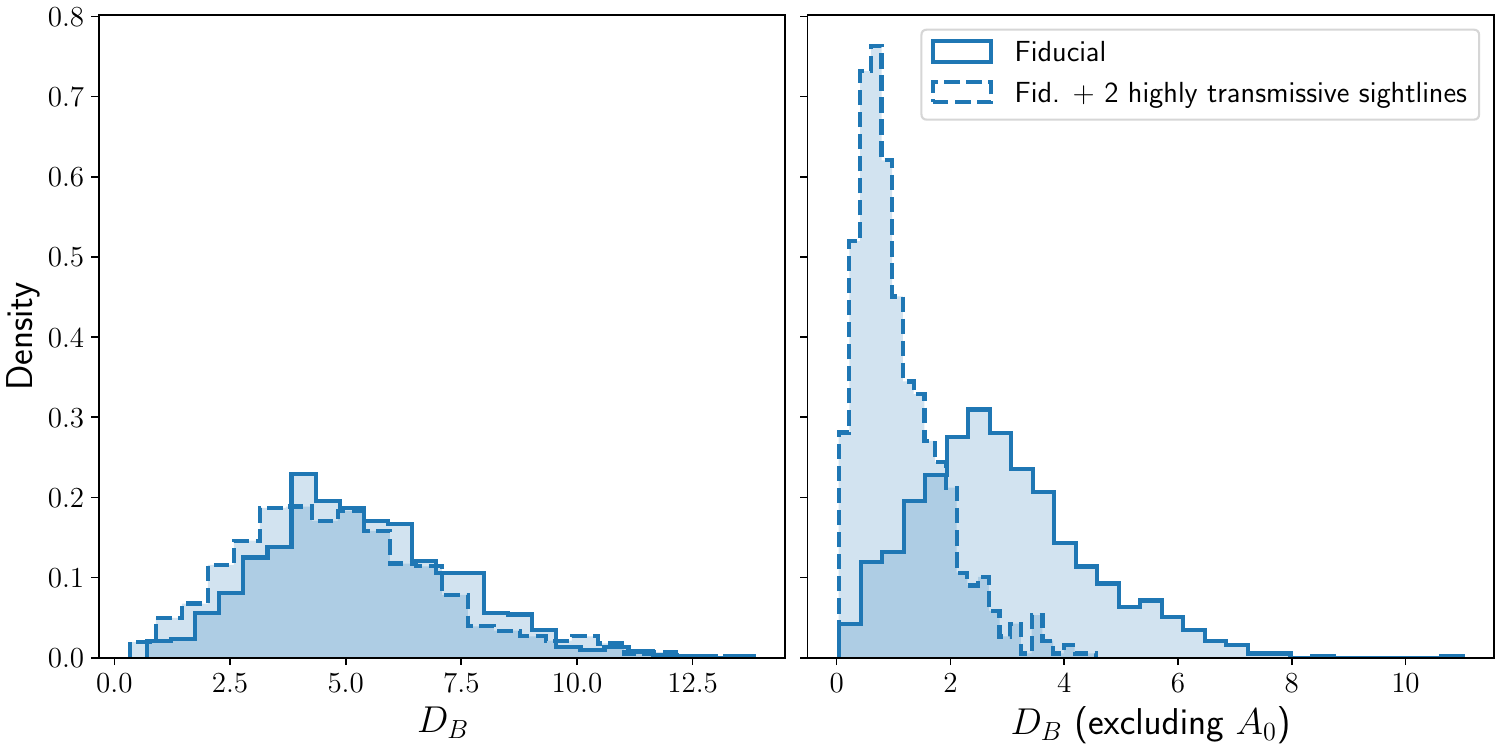}
\caption{Distribution of $D_B$ for mock realizations of the fiducial model (solid), and for the modified mock realizations in which two sightlines have been replaced to be highly transmissive at $z>5.5$ (dashed).  In the left-hand panel, $D_B$ is computed using all four correlation parameters, while in the right-hand panel the $A_0$ parameter is excluded. \label{fig:bhattacharyya}}
\end{figure}

Similar to section~\ref{subsec:fiducial_model}, we compute the Bhattacharyya distance between the posterior distributions of the modified mock realizations and that of the observed sample. The resulting distribution is shown in figure~\ref{fig:bhattacharyya} (left panel) with dashed lines. We find only a slight shift of the distribution toward lower distance values compared to the fiducial case. Using a $D_B$ threshold of 1.5, motivated by the clear separation between the most discrepant jackknife pair ($D_B=1.90$) and all remaining cases ($D_B<1.5$) in figure~\ref{fig:jackknife_corner}, we find that the fraction of realizations consistent with the observations increases to 4.1\%, compared to 2\% in the fiducial case. We emphasize that this threshold is intended as an empirical consistency criterion rather than a formal statistical significance threshold. However, as seen in figure~\ref{fig:correlations_biased_two}, a significant offset remains in the evolution of the correlation amplitude. Although the slope is broadly consistent, the value of the correlation amplitude at $z=5$ (corresponding to parameter $A_0$ in our model) inferred from the modified mock realizations differs substantially from that obtained from the observations.

A possible explanation for the remaining mismatch in $A_0$ is that the correlation amplitude is sensitive to systematic uncertainties in the continuum normalization of the observed spectra, which are not explicitly included in our mock realizations. Such effects would primarily introduce an overall shift in the normalization of the correlation matrix rather than a change in its redshift-dependent structure, and would therefore affect $A_0$ more strongly than the slope parameter $A_1$ or the inferred correlation length. Motivated by this, we recalculate the $D_B$ values after excluding $A_0$ from the parameter vector, thereby focusing on the relative redshift evolution and correlation length parameters that are expected to be less sensitive to continuum-normalization uncertainties. The resulting distribution is shown in figure~\ref{fig:bhattacharyya} (right panel). We find that $17.5\%$ of the fiducial realizations satisfy our consistency criterion, $D_B$ (excluding $A_0$) $<1.5$. Thus, the fiducial model naturally produces realizations consistent with the observed redshift evolution and correlation length, without requiring the insertion of highly transmissive sightlines. For the modified realizations containing two highly transmissive sightlines, this fraction increases to $74.1\%$.

We further find that $11.4\%$ of the total sightlines in the fiducial lightcone have a mean transmitted flux at $z>5.5$ greater than or equal to that of the more transmissive of the two observed sightlines. Thus, a random realization of 67 sightlines contains, on average, $7.6$ such highly transmissive sightlines, demonstrating that these sightlines are not underproduced by the fiducial model. This raises the question of why only $17.5\%$ of the fiducial realizations reproduce the observed signal if highly transmissive sightlines are already sufficiently abundant.

To investigate this, we compare the transmitted-flux distributions of the fiducial and modified realizations at $z>5.9$. We use the Gini coefficient to quantify the inequality of the transmitted-flux distribution and $P_{97}/P_{50}$ to characterize its high-transmission tail. The results are summarized in Table~\ref{tab:gini_percentile}. The 17.5\% fiducial realizations satisfying $D_B$ (excluding $A_0$) $<1.5$ have values similar to those of the modified realizations, whereas the remaining fiducial realizations show lower values for both statistics. The quoted standard deviations are the scatter among realizations and therefore represent the expected uncertainty for a single observational sample of 67 sightlines. Since this scatter is comparable to the differences between the mean values of the modified realizations and those inconsistent with the data, neither statistic provides strong discrimination between the two populations for a sample of the current size. Larger observational samples will therefore be required to determine whether the observed correlations can be attributed entirely to cosmic variance or instead indicate missing physical ingredients in the reionization model.

\begin{table}[htbp]
\centering
\begin{tabular}{lccc}
\hline
Mock Realizations & $N$ & Gini & $P_{97}/P_{50}$ \\
\hline
Fiducial ($D_B<1.5$) & 175 & $0.673\pm0.051$ & $14.6\pm8.2$ \\
Fiducial ($D_B\geq1.5$) & 825 & $0.652\pm0.048$ & $11.9\pm5.4$ \\
Fiducial + 2 highly transmissive sightlines & 1000 & $0.673\pm0.048$ & $15.2\pm7.4$ \\
\hline
\end{tabular}
\caption{Gini coefficient and $P_{97}/P_{50}$ for the fiducial and modified mock realizations, evaluated over $z>5.9$. For the fiducial realizations, $D_B$ is calculated after excluding $A_0$. The quoted uncertainties are the scatter among realizations.}\label{tab:gini_percentile}
\end{table}

Taken together, these results indicate that it is not simply the presence or abundance of highly transmissive sightlines that is important, but rather whether these sightlines are sufficiently extreme relative to the other sightlines in a given realization. This also explains the result of our injection experiment: the two inserted sightlines were selected from the top $1\%$ of the transmitted-flux distribution and therefore remain highly transmissive relative to the other sightlines, allowing a much larger fraction of the modified realizations to reproduce the observed signal.

\subsection{Spatially uniform \texorpdfstring{$\lambda_{\mathrm{mfp}}$}{ionizing mean free path} model} 

Before concluding, we examine the importance of one of the key ingredients of our fiducial model: spatial fluctuations in the ionizing mean free path, $\lambda_\mathrm{mfp}$. To assess their impact, we repeat our analysis using a modified model in which $\lambda_\mathrm{mfp}$ is assumed to be spatially uniform, while still evolving with redshift. Specifically, we ensure that the redshift evolution of the globally averaged $\lambda_\mathrm{mfp}$\footnote{The globally averaged mean free path is calculated using eq.~2.63 of ref.~\cite{choudhuryCapturingSmallscaleReionization2025}.} is identical to that in the fiducial model. As in the fiducial case, the $\kappa_{\mathrm{res}}$ values were tuned to reproduce the observed $\tau_{\mathrm{eff}}$ distribution, ensuring a fair comparison between the two models. All other model parameters are kept identical to those of the fiducial model. Comparing these two scenarios allows us to isolate the role of spatial $\lambda_\mathrm{mfp}$ fluctuations in reproducing the observed large-scale Ly$\alpha$ forest correlations.

\begin{figure}[htbp]
\centering
\includegraphics[width=0.75\textwidth]{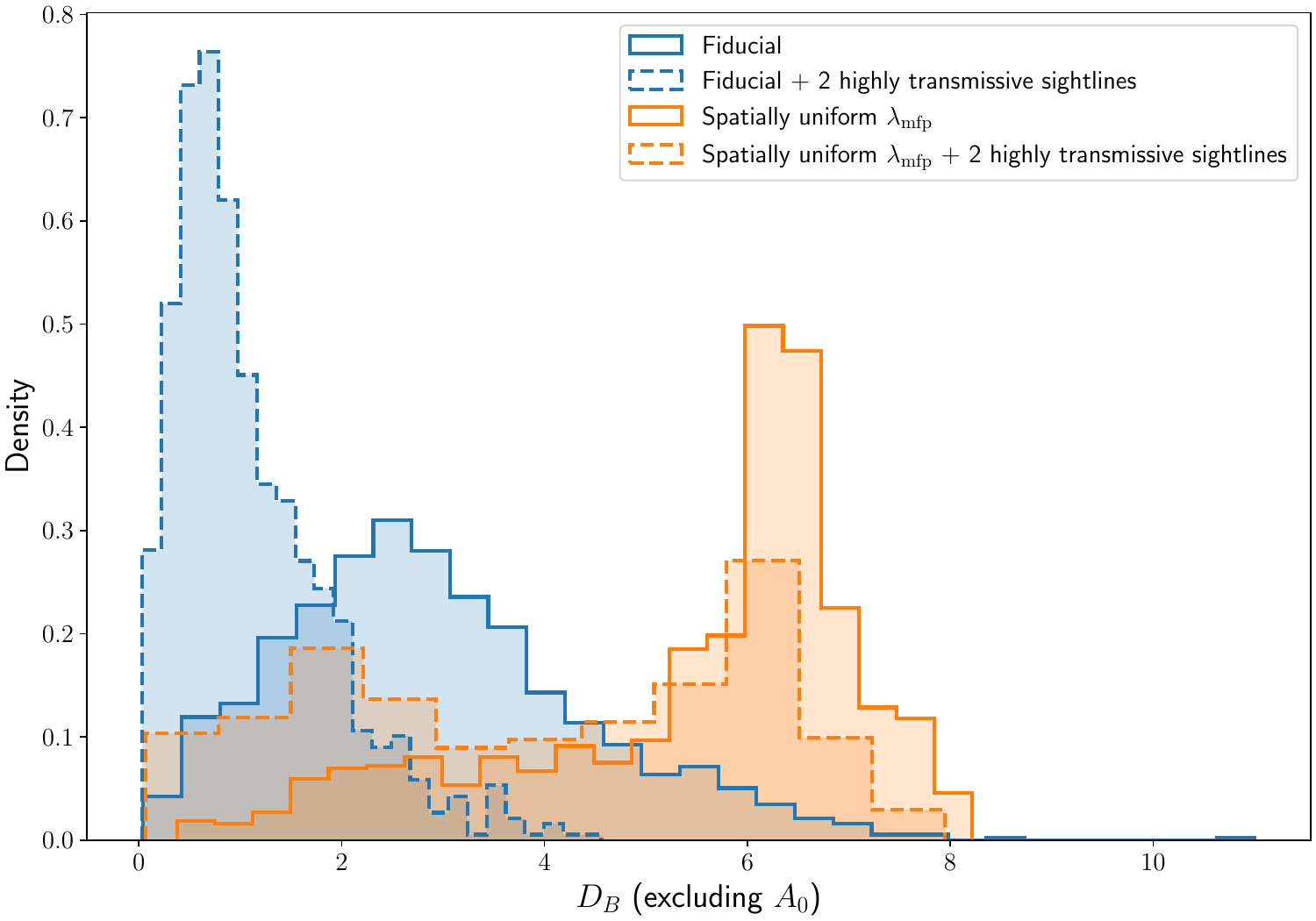}
\caption{Distribution of $D_B$ for mock realizations of the fiducial model with spatially fluctuating $\lambda_\mathrm{mfp}$ (blue) and the spatially uniform $\lambda_\mathrm{mfp}$ model (orange). The dashed histograms show the corresponding distributions after replacing two sightlines in each realization with sightlines that are highly transmissive at $z>5.5$. The parameter $A_0$ is excluded from the $D_B$ calculation. \label{fig:bhattacharyya_lin_mfp}}
\end{figure}

The resulting distributions of $D_B$ are shown in figure~\ref{fig:bhattacharyya_lin_mfp}. The model with spatially uniform $\lambda_\mathrm{mfp}$ produces systematically larger $D_B$ values than the fiducial model with fluctuating $\lambda_\mathrm{mfp}$, indicating significantly poorer agreement with the observed large-scale Ly$\alpha$ forest correlations. This demonstrates that spatial fluctuations in the ionizing mean free path constitute a crucial ingredient for reproducing the observed correlation signal. Physically, spatial variations in $\lambda_\mathrm{mfp}$ introduce additional large-scale fluctuations in the ionizing background, thereby enhancing coherence in the transmitted flux over large distances.

At the same time, both models exhibit a noticeable shift toward lower $D_B$ values when highly transmissive sightlines are included. This indicates that such rare sightlines can substantially strengthen the inferred correlation signal, even in models that otherwise provide a poorer match to the observations.

\section{Summary} \label{sec:summary}

Recent analyses of existing high-redshift Ly$\alpha$ spectra have brought a new theoretical challenge to light: strong correlations on scales exceeding $200\mathrm{\,cMpc}$ at $z = 6$. In this work, we investigated whether these large-scale correlations can be reproduced within the extended \texttt{SCRIPT} semi-numerical reionization model. We generated 1000 mock realizations of 67 quasar sightlines, carefully matching the redshift coverage, masking, and noise properties of the observational sample. Using a Markov chain Monte Carlo framework, we inferred the correlation amplitude and length parameters for both the observed data and our mock ensembles.  

We found that while the ensemble average of our fiducial model systematically predicts smaller correlation lengths than those inferred from the data, $17.5\%$ of individual realizations naturally reproduce the observed signal. To understand this, we performed a delete-2 jackknife analysis on the observational sample, which revealed that the inferred correlation length is disproportionately driven by a rare pair of highly transmissive sightlines (corresponding to quasars J1535$+$1943 and PSO J011$+$09). The high transmissivity of these sightlines is associated with strong high-redshift transmission spikes. 

Motivated by this finding, we modified our mock realizations by selectively inserting two skewers that are highly transmissive at $z > 5.5$. This modification significantly increased the fraction of mock realizations consistent with the observed redshift evolution and correlation length from 17.5\% in the fiducial case to 74.1\%. Furthermore, by comparing our results to an alternative model featuring a spatially uniform ionizing mean free path, $\lambda_{\mathrm{mfp}}$, we demonstrated that spatial fluctuations in $\lambda_{\mathrm{mfp}}$ remain an essential physical ingredient for reproducing the observed large-scale correlation structure.

Ultimately, our results suggest that the unexpectedly large Ly$\alpha$ correlations do not necessarily require physics beyond standard late-reionization models. Instead, they can be reconciled with existing frameworks when accounting for cosmic variance and the outsized statistical impact of highly transmissive sightlines that occur as prominent outliers within the observational sample. However, the present observational sample cannot conclusively distinguish an uncommon realization arising from cosmic variance from a scenario in which additional physical ingredients are required. Future observational samples containing substantially larger numbers of high-redshift quasars will provide a more decisive test.

\appendix

\section{Validation} 

\subsection{\texorpdfstring{$\tau_{\mathrm{eff}}$}{effective optical depth} distributions} \label{apdx:validation_taueffs}

In this appendix, we compare the distribution of effective Ly$\alpha$ forest optical depths obtained from our fiducial model with observations~\cite{bosmanHydrogenReionizationEnds2022}. This serves as a validation of the formalism used to compute the Ly$\alpha$ optical depth, as described in section~\ref{subsubsec:optical_depth}.

\begin{figure}[htbp]
\centering
\includegraphics[width=\textwidth]{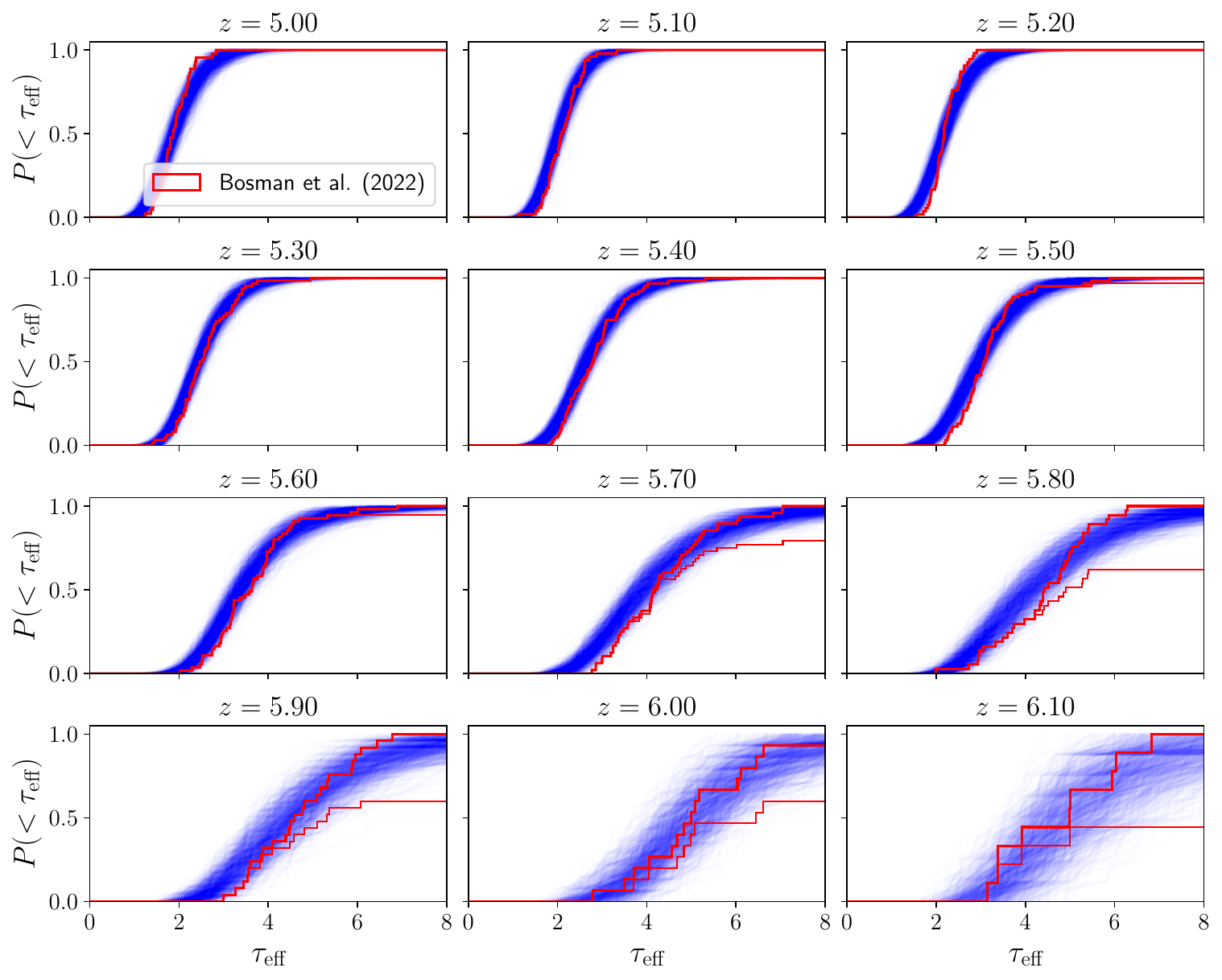}
\caption{Effective optical depth CDFs, averaged over redshift bins of width $\Delta z=0.1$, for the fiducial model. The red curves show the observational measurements from Bosman et al. (2022)~\cite{bosmanHydrogenReionizationEnds2022}. The upper red curves correspond to the case where lower limits are assigned values just below the detection threshold, while the lower red curves assume lower limits of $\tau_{\mathrm{eff}}\rightarrow\infty$. The blue curves indicate the model predictions obtained from 1000 mock realizations, each containing the same number of sightlines as the observational sample. \label{fig:tau_eff_validation}}
\end{figure}

To this end, we divide each skewer in our transmitted-flux lightcone into redshift bins of width $\Delta z = 0.1$. The effective Ly$\alpha$ optical depth for a given bin is then defined as
\begin{align}
\tau_{\mathrm{eff}} = -\ln\left(\frac{1}{n}\sum_i F_i\right),
\end{align}
where $F_i$ is given by eq.~\ref{eq:flux}, and the sum runs over all $n$ redshift points within the chosen bin. To mimic the observational sample, we generate multiple realizations by randomly selecting a number of sightlines equal to the number of observed sightlines in each redshift bin, and compute the cumulative distribution function (CDF) of $\tau_{\mathrm{eff}}$ for each realization. The resulting ensemble of CDFs for the fiducial model is compared with the observed distributions in figure~\ref{fig:tau_eff_validation}. 

\begin{figure}[htbp]
\centering
\includegraphics[width=\textwidth]{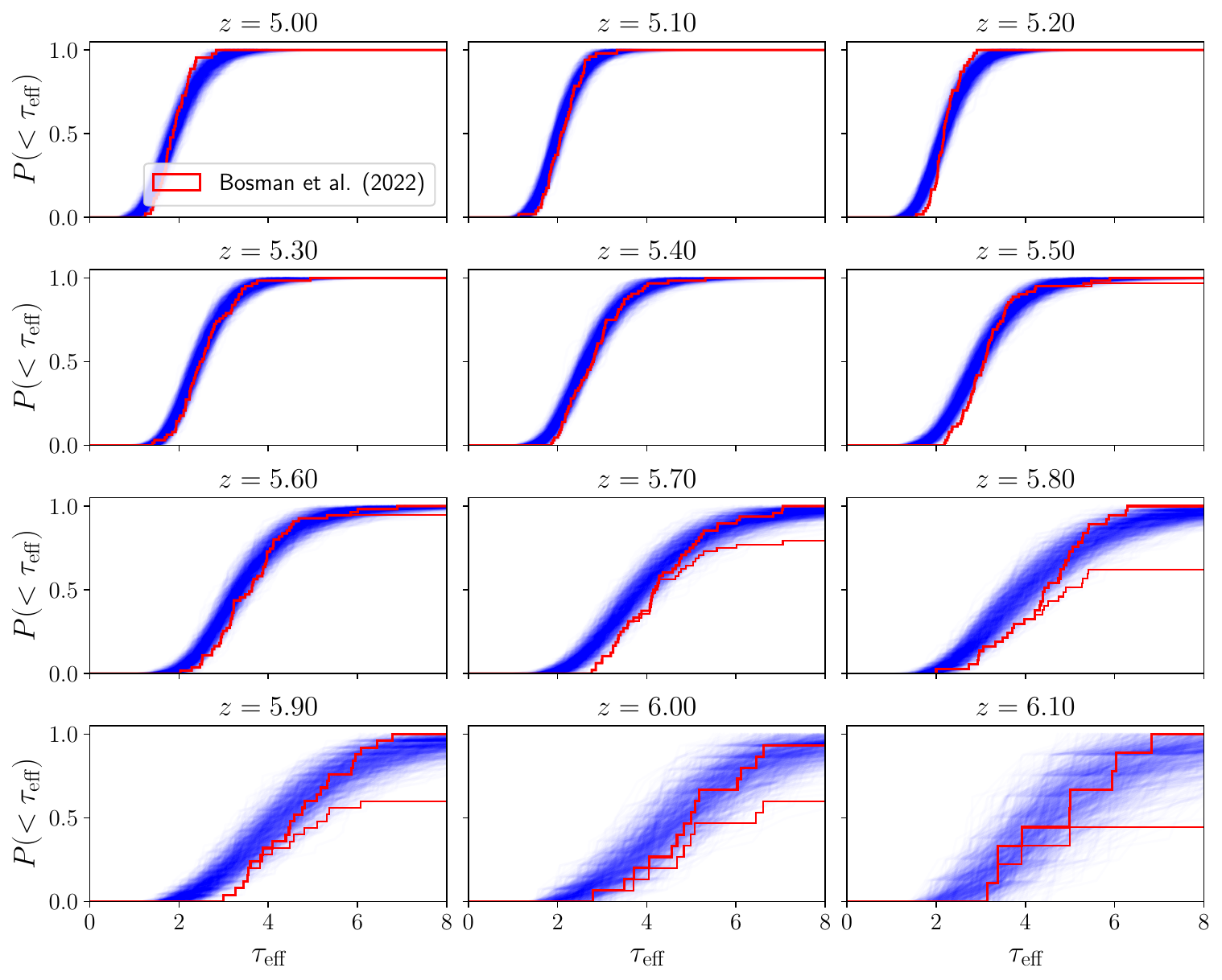}
\caption{Effective optical depth CDFs, averaged over redshift bins of width $\Delta z=0.1$, for the fiducial model with two highly transmissive sightlines inserted. Formatting is the same as figure~\ref{fig:tau_eff_validation}. \label{fig:tau_eff_validation_biased}}
\end{figure}

We also repeat the same validation analysis after inserting mock highly transmissive sightlines into those redshift bins in which such sightlines are present in the observational data. The resulting ensemble of CDFs is compared with the observed distributions in figure~\ref{fig:tau_eff_validation_biased}. As expected, the CDFs, especially for the high-redshift bins, shift slightly toward lower $\tau_{\mathrm{eff}}$ values, while remaining overall consistent with the observations.

\subsection{Analysis pipeline} \label{apdx:validation_analysis_pipeline}

In this appendix, we compare the correlation parameters inferred from the observational data used in this work with those reported in ref.~\cite{spinaMeasuringIntergalacticMedium2026}. This provides a validation of our correlation analysis pipeline. 

\begin{figure}[htbp]
\centering
\includegraphics[width=\textwidth]{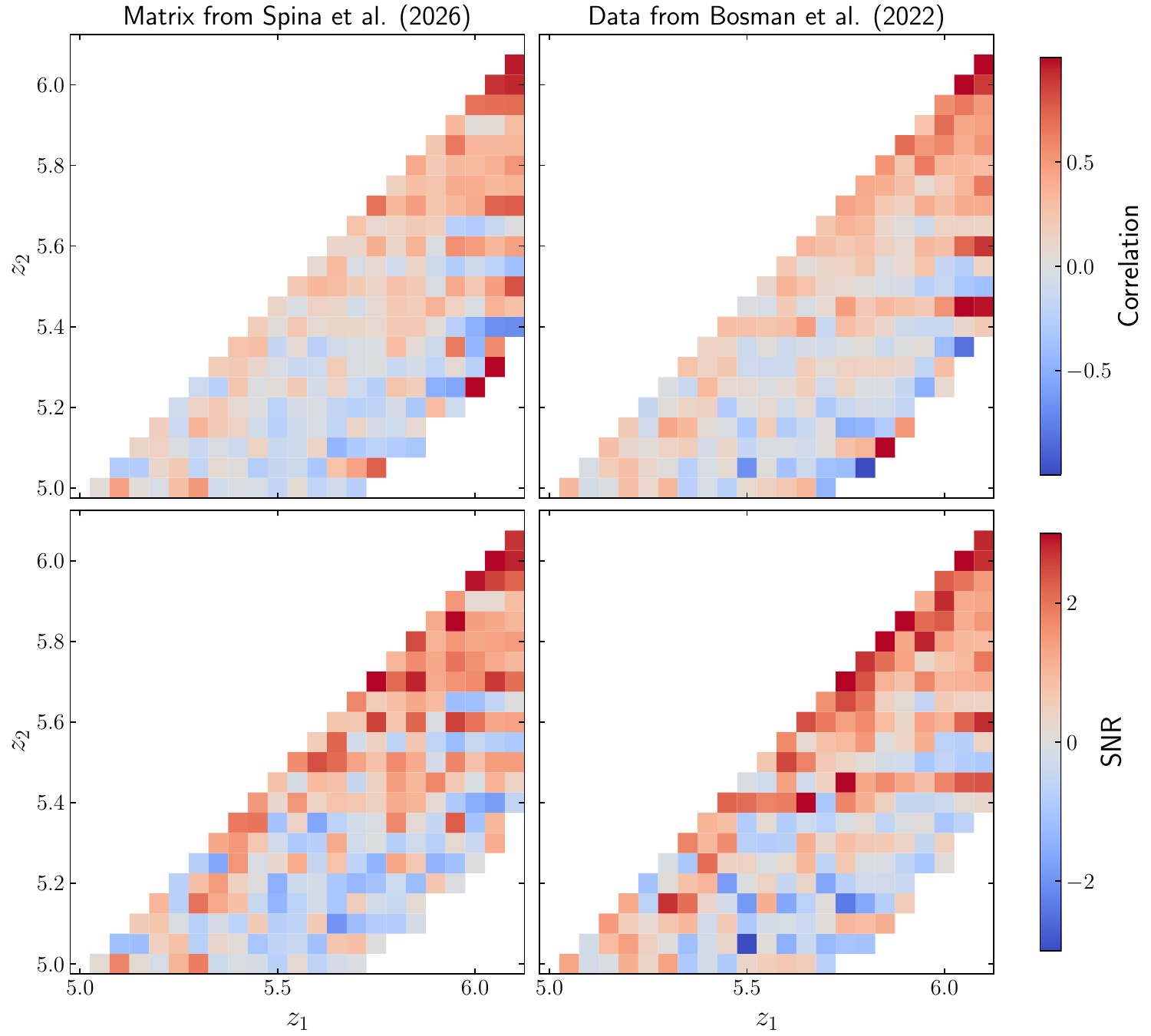}
\caption{Comparison of the published correlation matrix $\mathcal{D}_{jk}$, and the SNR matrix $\mathcal{D}_{jk}/\sigma_{jk}$ from Spina et al. (2026)~\cite{spinaMeasuringIntergalacticMedium2026} (\textit{left panel}) with the ones computed in this work using the observed data from Bosman et al. (2022)~\cite{bosmanHydrogenReionizationEnds2022} (\textit{right panel}). \label{fig:corr_matrices}}
\end{figure}

\begin{figure}[htbp]
\centering
\includegraphics[width=\textwidth]{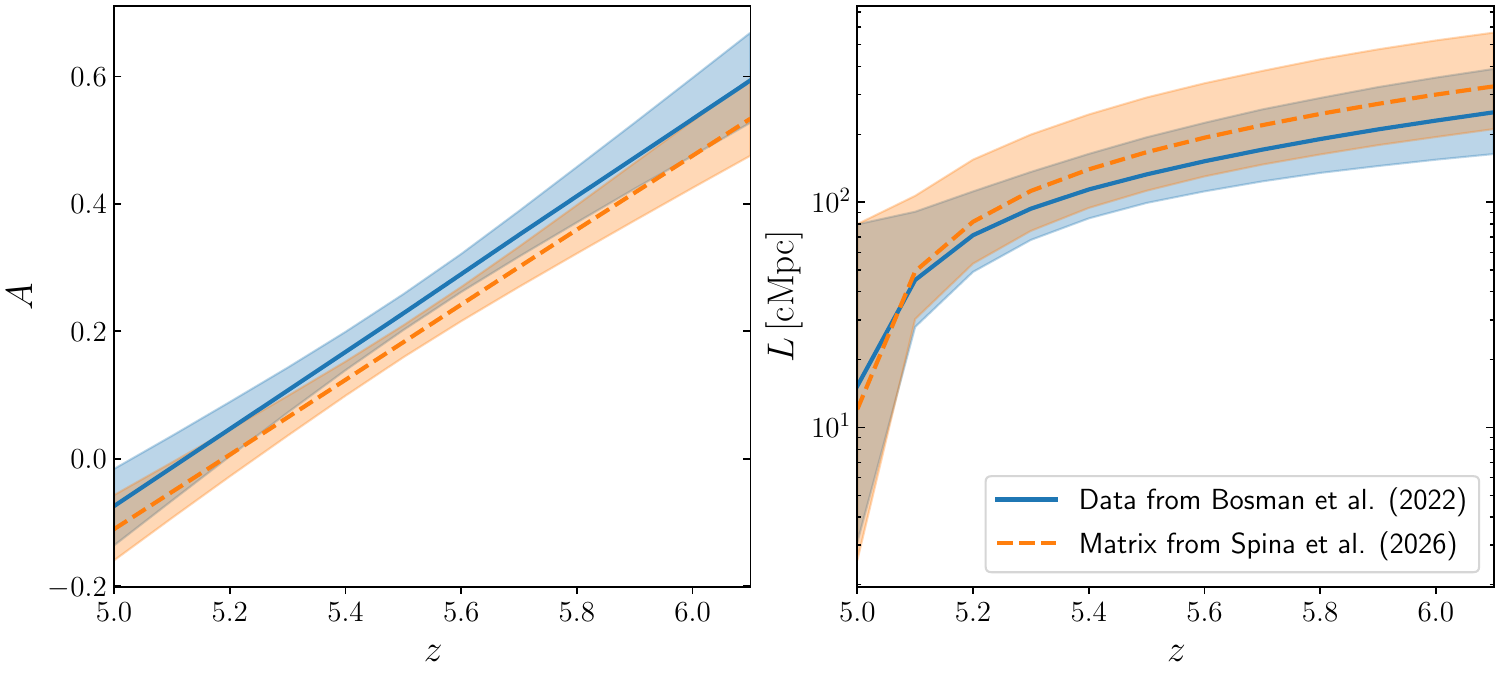}
\caption{Redshift evolution of the correlation amplitude and length. The results obtained using the observed data from
Bosman et al. (2022)~\cite{bosmanHydrogenReionizationEnds2022} are shown in blue. The orange curves show the corresponding results obtained using the published correlation matrix from
Spina et al. (2026)~\cite{spinaMeasuringIntergalacticMedium2026}. Formatting is the same as in figure~\ref{fig:correlations_fiducial}. \label{fig:validation}}
\end{figure}

To reproduce the correlation measurements reported in ref.~\cite{spinaMeasuringIntergalacticMedium2026}, we use their published correlation matrix\footnote{\url{https://cdsarc.cds.unistra.fr/viz-bin/cat/J/A+A/706/A273}} and perform the MCMC analysis described in section~\ref{subsec:modeling_inference}. We then compare this posterior distribution with the one obtained using the correlation matrix that we estimate from the observed spectra directly. In figure~\ref{fig:corr_matrices}, we compare the published correlation matrix $\mathcal{D}_{jk}$, and the SNR matrix $\mathcal{D}_{jk}/\sigma_{jk}$ with the ones computed from the observed spectra. Although they are not identical, owing to minor differences in the processing pipelines, they exhibit the same overall structure. Figure~\ref{fig:validation} compares the redshift evolution of the correlation amplitude and correlation length inferred from the two correlation matrices, showing good agreement and thereby validating our analysis pipeline.

\acknowledgments
The authors acknowledge support of the Department of Atomic Energy, Government of India, under project no. 12-R\&D-TFR-5.02-0700. The authors thank Ravi K. Sheth and Aseem Paranjape for an insightful discussion and helpful suggestions. We also thank the authors of Bosman et al. (2022) and Spina et al. (2026) for making the transmitted-flux measurements and the measured correlation matrix publicly available.

\paragraph{Data Availability Statement.} The data generated and presented in this paper will be made available upon reasonable request to the corresponding author.

\bibliographystyle{JHEP}
\bibliography{biblio.bib}

\end{document}